%% file: main.tex
\documentclass[journal]{IEEEtran}
\usepackage{amssymb, amsmath}
\usepackage{bm}
\usepackage{color}
\usepackage{graphicx}
\usepackage{cite}
\usepackage{booktabs}
\usepackage{multirow}
\usepackage{threeparttable}
\usepackage{enumerate}
\usepackage[table]{xcolor}
\usepackage[english]{babel}
\usepackage{algorithmicx}
\usepackage{algorithm}
\usepackage{algpseudocode}
\usepackage{nomencl}
\usepackage{multicol}
\usepackage{commath}
\usepackage{makecell}
\usepackage{blindtext}
\usepackage{soul,comment}
\usepackage[belowskip=-10pt,aboveskip=2pt]{caption}
\input{preamble.tex}
\makenomenclature
\graphicspath{{pics/}}

\begin{document}
\title{PIDGeuN: Graph Neural Network-Enabled Transient Dynamics Prediction of  Networked Microgrids Through Full-Field Measurement}
\author{Yin Yu, Xinyuan Jiang, Daning Huang, and Yan Li,~\IEEEmembership{Senior Member,~IEEE}
\thanks{Y. Yu and D. Huang are with the Department of Aerospace Engineering, The Pennsylvania State University, University Park, PA 16802, USA (e-mail: daning@psu.edu).  X. Jiang and Y. Li is with the Department of Electrical Engineering, The Pennsylvania State University, University Park, PA 16802, USA (e-mail: yql5925@psu.edu).}
}

\maketitle

\input{./src/abstract}
\input{./src/introduction}
\input{./src/formulation}
\input{./src/network}
\input{./src/numExample}

\input{./src/conclusion}

\bibliographystyle{IEEEtran}
\bibliography{main}

\end{document}

%% file: preamble.tex
\usepackage{new-mysymb}
\usepackage{longtable}
\usepackage{multirow}
\usepackage{pbox}
\usepackage[hyphens]{url}
\usepackage{xcolor}
\usepackage{soul}

\newcommand\bR{\mathbb{R}}

\newcommand\cG{\mathcal{G}}

\newcommand\cN{\mathcal{N}}

\newcommand\cW{\mathcal{W}}
\newcommand\cX{\mathcal{X}}
\newcommand\cV{\mathcal{V}}
\newcommand\cE{\mathcal{E}}

\newcommand\Dt{\Delta t}

\newcommand\dvX{\dot{\vX}}

\newcommand\tvL{\Tilde{\vL}}

\newcommand\cg{\cellcolor{gray!30}}

\soulregister\cite7
\soulregister\ref7

%% file: src/abstract.tex
\begin{abstract}

A Physics-Informed Dynamic Graph Neural Network (PIDGeuN) is presented to accurately, efficiently and robustly predict the nonlinear transient dynamics of microgrids in the presence of disturbances.
The graph-based architecture of PIDGeuN provides a natural representation of the microgrid topology.  Using only the state information that is practically measurable, PIDGeuN employs a time delay embedding formulation to fully reproduce the system dynamics, avoiding  the dependency of conventional methods on internal dynamic states such as controllers. 
Based on a judiciously designed message passing mechanism, the PIDGeuN incorporates two physics-informed techniques to improve its prediction performance, including a physics-data-infusion approach to determining the inter-dependencies between buses, and a loss term to respect the known physical law of the power system, i.e., the Kirchhoff's law, to ensure the feasibility of the model prediction.
Extensive tests show that PIDGeuN can provide accurate and robust prediction of transient dynamics for nonlinear microgrids over a long-term time period.  Therefore, the PIDGeuN offers a potent tool for the modeling of large scale networked microgrids (NMs), with potential applications to predictive or preventive control in real time applications for the stable and resilient operations of  NMs. 

\end{abstract}


%% file: src/introduction.tex
\section{Introduction}
\PARstart{M}{odernization} of  electric power grid is critical for improving the  system's resiliency and reducing power outages
, e.g., Manhattan blackout. To solve this problem, microgrids have been recognized as a promising archetype by integrating Distributed Energy Resources (DERs), such as wind and photovoltaic (PV). To further enhance the flexible and resilient operations of low- or medium-voltage distribution networks, networked microgrids (NMs) are currently under development. 
Since most  DERs are integrated into microgrids through power-electronic interfaces, the system’s inertia is significantly reduced~\cite{
li2021cyber}.
Consequently, microgrids and NMs are sensitive and vulnerable to disturbances such as PV fluctuations, leading to frequent transient dynamics.

Although extensive research effort has been made on the microgrids' transient behavior \cite{sobbouhi2021transient}, it is still a  challenge to study and stabilize the system's transients. 
First, detailed modeling is usually required for studying the transients; however, the wide integration of DERs results in a  high-dimensional  system,  increasing the difficulty of efficiently analyzing the transient behavior.
Second, microgrids  are typical nonlinear systems, which is an inherent feature stemming from  power loads and dynamics of DERs; and thus, the existing model may not be sufficiently accurate to fully represent the nonlinear dynamical system.
Third, the operations of microgrids or NMs keep changing due to the fluctuations of DERs and/or the changes of system topology caused by the join or disconnection of microgrids or DERs.
With the wide deployment of the advanced metering infrastructure (AMI), the nonlinear dynamical system  is more observable than ever before.
Hence, one inspiring solution of transient dynamics  is: \emph{
to develop a data-driven approach to precisely and efficiently model and predict the system's transient dynamics, so that predictive or preventive control can be performed  to stabilize microgrids and NMs.
}


There are several existing data-driven approaches to identify the transient dynamics model of a nonlinear system through its operating data, which can fall into two major categories~\cite{natke2022recent}, namely linear models  and data-driven nonlinear methods. 
First, the linear models are well-established and commonly-used system identification methods. They develop a high-dimensional linear system with input and output to approximate the original nonlinear system's dynamics near an equilibrium point \cite{pierre2012overview}. 
The linear methods are relatively easy to implement and guaranteed to converge given sufficient system responses. However, the linear system methods do not have the extrapolation ability due to their nature of local linearization; and thus, it is not suitable to directly apply them to identify microgrid systems that are typically nonlinear for the entire operating envelope.
Second, several data-driven methods have also been developed to identify a nonlinear system to capture the global transient dynamics  over the entire state space \cite{fu2013research}. 
Theoretically, these methods can identify an accurate model if appropriate  nonlinear terms are used.  However, the selection of  correct nonlinear terms is a nontrivial task and the required number of terms grows exponentially as the system size increases. Moreover, these methods usually involve system's state variables that are hard to measure in practice.   Therefore, general nonlinear system identification methods soon become intractable when applied to the identification of practical nonlinear systems such as  microgrids.

Microgrids can be defined on buses and their pairwise connections. Such connections are ignored in many data-driven modeling  approaches, yet the intrinsic network topology is worth exploiting, as the connections partially govern its transient dynamics.  
The Graph Neural Network (GNN), 
a recent variant of deep learning models, has emerged as a powerful tool for the modeling of data defined on graphs, and thus a promising candidate for the transient dynamics modeling of microgrids.
In general, the GNN-based methods have gained tractions for many complex dynamical physics simulations that can benefit from the graph representation of the underlying systems, e.g., N-body dynamics \cite{Battaglia2016}, Hamiltonian mechanics \cite{Sanchez2019}, and particle dynamics \cite{SG2020}.
Furthermore, to tackle the  temporal dependency of  the transient problems, the spatial-temporal GNN (STGNN) is developed, where a form of recurrent architecture is added to the basic GNN to capture the temporal effects from the data.  The STGNN methods have shown superior performance over many conventional non-graph-based methods \cite{Seo2018,Chen2018}, and been successfully applied to the traffic flow prediction \cite{Bing2018,Guo2019,Zhao2020}.

The superior modeling capabilities of GNN has incurred interests in its application to power systems including microgrids, that have a natural graph structure. Most of the GNN-oriented studies on  power systems focus on static problems, e.g. optimal power flow (OPF) problem \cite{Owerko2020}, power flow approximations \cite{Bolz2019,Donon2019,Jeddi2021},  state estimation \cite{Hossain2021,Kundacina2022,Pagnier2021}, and anomaly detection 
\cite{Owerko2018,Fan2020,ML2021}.
For these static problems, the scalable nature of GNN makes it powerful to handle large systems in an efficient manner.  For example, in \cite{Owerko2020}, where the OPF problem is solved on IEEE-30 and IEEE-118 test cases, 
the conventional optimization-based methods saw a 8-fold increase in computational time on the larger grid, whereas the GNN-based methods only required $9\%$ more computational time on average. The model performance can be further improved when the known knowledge of the power system is infused into the network training. In \cite{Bolz2019} and \cite{Jeddi2021}, the authors applied the Kirchhoff's law as the training objective for their models in the power flow approximations to ensure that the network predictions are physically feasible.
Fewer efforts based on the GNN methods have been devoted to the modeling of transient dynamics in power systems, e.g. in the short-term power prediction of DERs 
\cite{Karimi2021,Jiao2021,Khodayar2021,Simeunovic2021},
or transient stability assessment (TSA) \cite{Sun2022,Qiao2021,Zhou2021,Nauck2022}.  However, these problems are considered on a large, relatively slow timescale, on which the transient responses 
in  microgrids are not well resolved. 
Therefore, in the literature, it still remains an open question whether a GNN-based data-driven model can be developed to capture and resolve the transient dynamics of a power system, esp. the NMs.

To bridge the gap identified above, we develop a novel 
Physics-Informed Dynamic Graph Neural Network (PIDGeuN), which is  a data-driven approach for accurate, efficient, robust and time-resolved prediction of microgrid transient dynamics.
The PIDGeuN incorporates the underlying physical laws and knowledge that the dynamic power system shall comply with in the stages of training and prediction in two aspects, with the following motivations and novelties.
\textit{First}, the strength of correlation between  buses, characterized by the pairwise edge weights in the graph, are typically fixed and predefined using the nodal admittance matrix; however, such weights are not necessarily good representations of the inter-dependencies between buses, since they are also affected by other factors, e.g. bus types. To best inform the prediction of the transient dynamics, our PIDGeuN is formulated to learn and dynamically adjust the edge weights based on the bus states and admittance matrix.
\textit{Second}, in many time series prediction problems, the prediction error accumulates over time and thus the long-term forecasting becomes a challenge; in the context of microgrids, the prediction error is likely to manifest in the form of nonphysical loss or gain of conserved energy, i.e., changes in electrical power that does not comply with the Kirchhoff’s law.  
Therefore to enable the stable long-term prediction of the data-driven model, the compliance with the power flow equation is explicitly enforced in the training of the PIDGeuN model.
The incorporation of the two physics-informed techniques improves the network's training efficiency and prediction accuracy, and enables the unprecedented capability of accurate long-term prediction for microgrid transient dynamics.  Such capability paves the way for performing  real-time control  in microgrids, which is the authors' next work.


The remainder of this paper is organized as follows: 
Section II poses the mathematical problem for the graph-based dynamic modeling of networked microgrids. 
Section  III introduces the presented PIDGeuN method for transient dynamics prediction.
In Section IV, the PIDGeuN model is benchmarked against existing methods using  numerical examples to demonstrate the feasibility and effectiveness of PIDGeuN.
Conclusions are drawn in Section V.

%% file: src/formulation.tex
\section{Problem Statement}\label{sec:formulation}

\subsection{Networked Microgrid Systems}
Networked microgrids consist of several distributed and independent microgrids to provide local energy generation and delivery.
Each microgrid is a group of  DERs and loads within clearly defined electrical boundaries, which acts as a single controllable
entity and can connect to or disconnect from NMs. 
Assume in a NM system, there are $\mathbb{G}$ DERs, $\mathbb{L}$ power loads, and $N$ buses $\{B_i\}_{i=1}^{N}$. The connection of buses is depicted by the admittance matrix $Y\in\mathbb{C}^{N\times N}$.
Each $B_i$ are described by the following quantities: 
active power $P_i\in\mathbb{R}$, 
reactive power $Q_i\in\mathbb{R}$, 
voltage $V_i\angle\delta_i\in\mathbb{C}$, and 
current $I_i\angle\theta_i\in\mathbb{C}$, which are measurable. 
These variables define a vector $\vx_{gi}\in\bR^6$,
\begin{equation}
    \vx_{gi} = [P_i,Q_i,\re{V_i\angle\delta_i},\im{V_i\angle\delta_i},\re{I_i\angle\theta_i},\im{I_i\angle\theta_i}]
\end{equation}
Note that the dynamics of microgrids are determined by several factors such as controller of DERs, power loads, network topology, etc.. In this work, only the measurable variables such as $\{\vx_{gi}\}_{i=1}^N$ are utilized  to identify the system dynamics. It removes the dependence of conventional data-driven methods on the internal states of DERs that are hard to measure,  making it feasible for real world applications.


At the steady state, DERs produce power to  satisfy the consumption of power loads. When disturbance occurs, the outputs of  dispatchable DERs are adjusted accordingly as well as power loads to compensate for the disturbance until a new equilibrium is reached. Our goal is to accurately predict the transients of the system in between two equilibrium points. 
 
Assuming buses are measured locally, we collect the measurements  within a period of time $T$ to identify the dynamical system.
To model the transient dynamics of  microgrids subject to disturbances, 
the bus $B_i$ at the time instance $k$ is then characterized by an extended state vector,
\begin{equation}
    \vx_i^{(k)}=[\vx_{gi},dP_i,dQ_i,\gamma,\beta_i]^{(k)}\in\bR^{10},
\end{equation}
where the first six variables correspond to the standard states $\vx_{gi}$. The new variables are introduced below.

The power disturbances are parametrized by $dP$ and $dQ$,
\begin{subequations}
\begin{align}\label{eqn:dp}
    dP_i^{(k)} &= P_i^{(k+1)} - P_i^{(k)}, \\\label{eqn:dq}
    dQ_i^{(k)} &= Q_i^{(k+1)} - Q_i^{(k)},
\end{align}
\end{subequations}
and the values of $dP$ and $dQ$ are non-zero \textit{only} in the buses where the disturbance occurs such as load or DER buses. 
A Boolean variable $\gamma$ is introduced to indicate if disturbances occur in \textit{any} of the buses, i.e., $\gamma^{(k)}=1$ means a disturbance occurred in the system at the $k^\text{th}$ time step (though not necessarily at bus $B_i$), and $\gamma^{(k)}=0$ otherwise. 
$\beta_i$ is the type index, meaning the type of the bus $B_i$, where,
\begin{equation}\label{eqn:busType}
    \beta_{i}=\left\{
    \begin{array}{ll}
        0 & \mbox{Empty} \\
        0.5 & \mbox{Loads} \\
        1.0 & \mbox{DERs (w/ or w/o Loads)}
    \end{array}\right.
\end{equation} 

At the time step $t=t_k$, denote the collection of the extended states as $\vX^{(k)}=\{\vx_i^{(k)}\}_{i=1}^N\in\bR^{N\times 10}$ and the standard states as $\vX_g^{(k)}=\{\vx_{gi}^{(k)}\}_{i=1}^N\in\bR^{N\times 6}$.  
The transient dynamics modeling and prediction of microgrids is stated as follows: Given a sequence of states of $C$ steps, $\cX_C^{(k)} = [\vX^{(k)},\vX^{(k-1)},\cdots,\vX^{(k-C+1)}]\in\bR^{N\times 10\times C}$, 
predict the system states $\vX_g^{(k+1)}$ at the future time $t=t_{k+1}$. Motivated by the time-delayed embedding technique \cite{Takens1981}, the use of consecutive time steps compensates for the partial knowledge of  microgrids obtained through AMI  and is necessary for the complete reconstruction of the microgrid dynamics.

\subsection{Graph Representation}
Power systems including microgrids can be represented by a graph, where the nodes are the buses of microgrids and the edges are the connections between buses. The weights of edges, i.e. the edge attributes, describe how correlated  the states $\vx_{gi}$ of two buses are. 
Formally, let $\mathcal{G}=(\mathcal{V},\mathcal{E},\mathcal{W})$ be a graph with a set of $N$ nodes, $\mathcal{V}$, a set of edges $\mathcal{E}\subseteq\mathcal{V}\times\mathcal{V}$ and the edge weights $\mathcal{W}$. The graph for microgrids is undirected, meaning that if $(i,j)\in\mathcal{E}$ then $(j,i)\in\mathcal{E}$.  Conventionally the edge weights are computed as given in (\ref{eqn:weights})~\cite{Owerko2020},
\begin{equation}\label{eqn:weights}
    w_{ij}=\left\{
    \begin{array}{ll}
        \exp\left(-k|Y_{ij}|^2\right) & (i,j)\in\mathcal{E} \\
        0 & \mbox{otherwise},
    \end{array}\right.
\end{equation} 
where $k$ is a scaling parameter that is tuned so that the weights are uniformly distributed.
Since the admittance matrix $Y$ is symmetric, $w_{ij}=w_{ji}$. The adjacency matrix of a graph that describes the connections between the nodes is defined by $\vA\in\mathbb{R}^{N\times N}$ where $[\vA]_{ij}=w_{ij}$ if $(i,j)\in\mathcal{E}$ and 0 otherwise. Therefore,  the diagonal terms in the adjacency matrix $\vA$ are always zero, whereas those in the admittance matrix $\vY$ are not. The graph is more conveniently represented by a normalized Laplacian matrix $\vL=\vI-\vD^{-1/2}\vA\vD^{-1/2}$,
where $\vD$ is a diagonal degree matrix with $[\vD]_{ii}=\sum_j[\vA]_{ij}$. The eigendecomposition of the Laplacian matrix, $\vL=\vQ\vtL\vQ^T$, defines the Graph Fourier Transform (GFT) \cite{Bruna2013}, a basic processing technique for data on graph, where the eigenvalues $\vtL$ represent frequencies and eigenvectors $\vtQ$ form the graph Fourier basis.

Data defined on a graph reside on a non-Euclidean space and often comes with a variable size of unordered nodes without a fixed spatial locality. The learning task on graph therefore poses challenges to the conventional machine learning algorithms that mostly work with matrices and multi-dimensional arrays, and requires the specially designed class of machine learning methods.

Utilizing GNNs, microgrid transient dynamics  is characterized by the following differential equation,
\begin{equation}\label{eqn:output}
    \dvX_g^{(k)}=\vF(\cX_C^{(k)},\cG;\vtQ),
\end{equation}
where a GNN $\vF$, parametrized by $\vtQ$, maps the sequence of $C$ consecutive extended state vectors $\cX_C^{(k)}$ to the rate of change $\dvX_g^{(k)}$ for the states on all buses at the current time step $k$, given the graph structure $\cG$ of the  system.
Then the states of the buses at the next time step can be numerically calculated. (\ref{eqn:nextState}) gives an example  when the explicit integration is adopted. 
\begin{equation}\label{eqn:nextState}
    \vX_g^{(k+1)}=\vX_g^{(k)}+\Dt\dvX_g^{(k)},
\end{equation}
where $\Dt$ is the time step size.

%% file: src/network.tex
\section{Formulation of PIDGeuN}\label{sec:network}

In this section, we present the key components and salient features of the proposed PIDGeuN architecture.  The architecture builds upon the message passing (MP) mechanism of GNNs and judiciously chooses a hybrid form of two MP implementations as its building blocks to capture the dynamical system's transient dynamics.  Furthermore, the known physical knowledge of microgrids is infused into both network architecture and loss function, in order to improve the expressiveness and training efficiency of the network.

\subsection{Message Passing Mechanism}\label{sec:mp}
The message passing mechanism is the corner stone for many GNN architectures, which consists of multiple consecutive MP steps.
Consider an input graph $\cG=(\cV,\cE,\cW)$ of $N$ nodes, and each node $v\in\cV$ has a node feature vector $\vh_v\in\mathbb{R}^D$ and a set of neighbor nodes $u\in\cN(v)$. At the $j^\text{th}$ MP step, the new feature of node $v$ is computed using its previous feature and information from its neighbors as \cite{Hamilton2020},
\begin{subequations}
\begin{align}\label{eqn:ebd}
  \vm^j_{\cN(v)} &= \textrm{AGGREGATE}\left(\{\vh_u^j\ |\ u\in\cN(v)\},\cW\right), \\
  \vh_v^{j+1} &= \textrm{UPDATE}\left(\vh_v^j,\vm^j_{\cN(v)},\cW\right),
\end{align}
\end{subequations}
where $\textrm{AGGREGATE}$ and $\textrm{UPDATE}$ are nonlinear mappings, e.g., neural networks, and $\vm_{\cN(v)}$ denotes the information aggregated from the neighbors of node $v$. 
One MP step corresponds to the information exchange between 1-hop neighbors, i.e., the nodes that directly connected. It is possible to stack different forms of aggregators over $k$ MP steps, and the feature vector of a node is influenced not only by its 1-hop neighbors, but also by the more distant $k$-hop neighbors.

For microgrids, using the MP mechanism, the network can predict the change of states of each node, determined by its input states and those of its neighbors through a sequence of neural network modules; the long-range interaction between the buses during a  disturbance is captured via a stack of MP layers.

\subsection{Two Typical MP Implementations}
\subsubsection{Graph Convolutional Layers (GCLs)}
The GCLs generalize the convolution operation from multi-dimensional data arrays to data on irregular topology, i.e., on graphs, and serve as an effective filter that extracts localized features from graph data.
Under the framework of GFT, the GCLs essentially performs the filtering on the spectrum of the Laplacian matrix \cite{Bruna2013}.
An efficient implementation of the GCL is the ChebConv network \cite{Defferrard2017}.  It avoids the direct computation of GFT and approximates the spectral filtering by a truncated series of Chebyshev polynomials $T_k$ up to $K^{\text{th}}$ order, 
which are equivalent to performing $K$ MP steps.

In ChebConv, the MP aggregation and updating is performed over all nodes simultaneously.
Let the input be $\vH^j=\{\vh_i^j\}_{i=1}^N\in\bR^{N\times D}$, the ChebConv-based graph convolution is defined as
\begin{equation}\label{eqn:cheb1}
    \vH^{j+1} = \sigma\left( \sum_{k=0}^{K}T_k(\Tilde{\vL})\vH^j\vtQ_k^j \right),
\end{equation}
where $\sigma$ is a nonlinear activation function, $\{\vtQ_k^j\}_{k=0}^K$ are learnable parameters, and the product $T_k(\tvL)\vH$ is computed recursively using
\begin{subequations}\label{eqn:cheb2}
\begin{align}
    T_0(\tvL)\vH &= \vH,\quad T_1(\tvL)\vH = \tvL\vH, \\
    T_k(\tvL)\vH &= 2\tvL T_{k-1}(\tvL)\vH-T_{k-2}(\tvL)\vH.
\end{align}
\end{subequations}
The ChebConv utilizes a scaled Laplacian $\tvL = (2/\lambda_{\max})\vL-\vI$ where $\lambda_{\max}$ is the largest eigenvalue of $\vL$, so that the eigenvalues of $\tvL$ range from $-1$ to $1$.  When compared to $\vL$, $\tvL$ promotes a more balanced contribution of the feature vectors from all the nodes to the output, esp. in irregular graphs \cite{Chung1997}.

\subsubsection{Graph Attention Layers (GALs)} The GAL can be viewed as a nonlinear form of graph convolution where the adjacency matrix is dynamically adjusted according to the node features using the attention mechanism \cite{Velickovic2018}. 
One GAL effectively performs just one MP step, but introduces stronger nonlinearity when compared to GCL.

In the general ``multi-head'' attention mechanism, for node $u$ and its neighbor $v$, with feature vectors $\vh_u^j$ and $\vh_v^j$, one may compute $m$ attention coefficients $\{\alpha_{uv}^{j,k}\}_{k=1}^m$ as
\begin{equation}\label{eqn:alpha}
    \alpha_{uv}^{j,k}=\frac{\exp(\textrm{LeakyReLU}(\vf_k(\vh_u^j,\vh_v^j;\vtQ_f^{j,k})))}{\sum_{w\in\cN(u)}\exp(\textrm{LeakyReLU}(\vf_k(\vh_u^j,\vh_w^j;\vtQ_f^{j,k})))}.
\end{equation}
In the attention calculation   (\ref{eqn:alpha}), $\vf_k$ is nonlinear function, such as a neural network, which characterizes the correlation between two feature vectors; LeakyReLU is a nonlinear activation function $f(x)=\max(-\epsilon x,x)$, where $\epsilon=0.2$ as a typical choice; and the sum-of-exp formulation normalizes the correlation to produce $\alpha_{uv}^j\in[0,1]$.  Subsequently, defining a set of new adjacency matrices $\{\vA_\alpha^{j,k}\}_{k=1}^m$, $[\vA_\alpha^{j,k}]_{uv}=\alpha_{uv}^{j,k}$, the node features are updated as
\begin{equation}\label{eqn:gal}
    \vH^{j+1} = \left\{\sigma( \vA_\alpha^{j,k}\vH^j\vtQ_\alpha^{j,k} )\right\}_{k=1}^m\in\bR^{N\times mD}.
\end{equation}
Note that in general $\vA_\alpha^{j,k}$ is not symmetric.  It is possible that a node $u$ is strongly influenced by its neighbor $v$, quantified by a large $\alpha_{uv}$, but not vice versa.

The multi-head attention mechanism is found to enhance the stability and expressive capability of the network as $m$ increases \cite{Wu2019}.  However, to reduce the computational cost and the model size, in this work an averaged adjacency matrix $\vA_\alpha^{j}$ is employed, $[\vA_\alpha^j]_{uv}=\frac{1}{m}\sum_{k=1}^m\alpha_{uv}^{j,k}$, and   (\ref{eqn:gal}) becomes $\vH^{j+1} = \sigma\left( \vA_\alpha^j\vH^j\vtQ_\alpha^j \right) \in\bR^{N\times D}$.  Overall the learnable parameters of a GAL include $\{\vtQ_\alpha^j\}\cap\{\vtQ_f^{j,k}\}_{k=1}^m$.  




\subsection{PIDGeuN Methodology}
\begin{figure}
    \centering
    \includegraphics[width=0.5\textwidth]{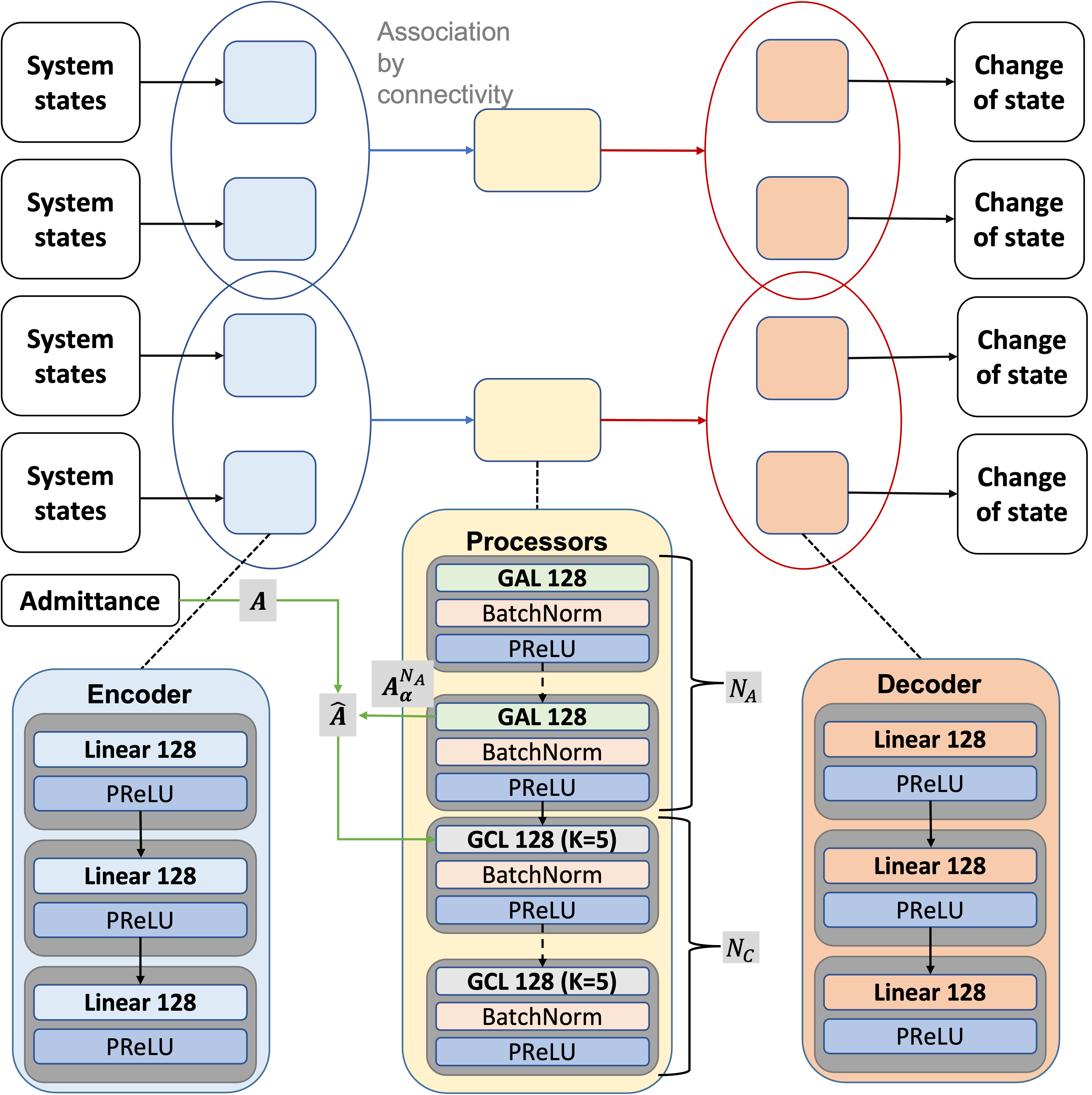}
    \caption{The PIDGeuN architecture.}
    \label{fig:networkArch}
\end{figure}

The PIDGeuN architecture, shown in Fig. \ref{fig:networkArch}, uses an encoder-processor-decoder architecture and adopts a combination of graph convolutional and attention layers with physics-informed techniques.
The key components of the PIDGeuN architecture are detailed in the following.


\subsubsection{Encoder} First, the encoder is applied to each individual node.  It maps microgrid state vectors at a node $\vx_i$, which consists of both continuous and discrete variables, to a latent vector $\vh_i^0\in\bR^D$.  The latent vector is a set of high-dimensional nonlinear features that provide a continuous representation of the states on each bus, which is amenable for NN computations. For the $i^{\text{th}}$ node at time step $k$, the encoder $\vf_E$ is
\begin{equation}\label{eqn:encoder}
    \vh_i^0 = \vf_E(\vx_i^{(k)},\vx_i^{(k-1)},\cdots,\vx_i^{(k-C+1)};\vtQ^0),
\end{equation}
where $\vf_E$ is implemented as a standard fully-connected NN (FCNN) of $N_M$ layers with a set of trainable parameters $\vtQ^0$. After the encoding, the latent vectors of all the nodes are denoted $\vH^0=\{\vh_i^0\}_{i=1}^N\in\bR^{N\times D}$.

\subsubsection{Processor} Subsequently, a stack of $N=N_A+N_C$ graph MP layers serve as processors that successively aggregate the latent features from each node and its neighbors and update the latent vectors at each node.  Formally, the $j^{\text{th}}$ processor step is written as
\begin{equation}
    \vH^{j+1}=\vf_P^j(\vH^j;\vtQ^j),
\end{equation}
where $\vf_P^j$ is either a GCL or a GAL, with parameter $\vtQ^j$.


Specifically, starting from the encoded latent vector $\vH^0$, the PIDGeuN first uses $N_A$ GAL layers in the processor to successively generate a series of latent vectors $\vH^1,\cdots,\vH^{N_A}$, as well as the attention-based adjacency matrix $\vA_\alpha^{N_A}$, using  (\ref{eqn:alpha}) and (\ref{eqn:gal}).

Next, to incorporate the physical knowledge of the microgrid into the network, a new \textit{physics-infused adjacency matrix} $\hat\vA$ is formed by combining the attention-based matrix $\vA_\alpha^{N_A}$ and the admittance-based matrix $\vA$ in  (\ref{eqn:weights}),
\begin{equation}
    \hat\vA = \frac{1}{2} \left( \vA_\alpha^{N_A}+\vA \right)
\end{equation}
Note that $\hat{\vA}$ maintains the same graph topology as the admittance-based $\vA$ in   (\ref{eqn:weights}), but with different non-symmetric weights.

The processing step is finalized with $N_C$ $K^{\text{th}}$-order GCL layers that use the normalized Laplacian $\hat{\vL}$ computed from $\hat{\vA}$, and generate a series of the latent vectors $\vH^{N_A+1},\cdots,\vH^{N}$ using   (\ref{eqn:cheb1}) and (\ref{eqn:cheb2}).  The last output $\vH^{N}$ is sent to the subsequent decoding step.

Over the entire processing step, the total number of effective MP step performed is $N_{MP} = N_A + KN_C$.
In the special case that $N_A=0$, the GCLs directly employ the symmetric admittance-based $\vA$ as the adjacency matrix; while when $N_C=0$, the GALs outputs the last latent vector $\vH^{N_A}$ for the next step and the attention coefficients are not used.

\subsubsection{Decoder} Finally, the decoder maps the latent vector of each node to the desired output, i.e. the rate of change,
\begin{equation}\label{eqn:decoder}
    \tilde{\dvX}_g^{(k)}=\vf_D(\vH^N;\vtQ^{N+1}),
\end{equation}
where $\vf_D$ is a FCNN of $N_M$ layers with trainable parameters $\vtQ^{N+1}$.

\subsubsection{Loss Function} The network parameters $\vtQ$  need to be trained using a loss function, a typically choice of which is the Frobenius-norm between the predicted and true rate of change over the training sequence of $N_t$ steps,
\begin{equation}\label{eqn:PredLoss}
    L_1(\vtQ)=\sum_{k=1}^{N_t}\norm{\tilde{\dvX}_g^{(k)}(\vtQ)-\dvX_g^{(k)}}_F^2
\end{equation}
However, leveraging known physical principles, at each time step the predicted states of the  grid, computed using   (\ref{eqn:nextState}), should satisfy the \textit{physical constraints}, i.e. the Kirchhoff's law, at each node, or power flow computation for the system. This fact motivates the inclusion of an additional term in the loss function to penalize the violation of the Kirchhoff's law in the prediction at each node and at each time step,
\begin{equation}\label{eqn:PILoss}
    L_2(\vtQ)=\sum_{i=1}^{N_k}\sum_{k=1}^{N_t}\lvert\tilde{P}_i^{(k)}+i\tilde{Q}_i^{(k)} - (\tilde{V}_i^{(k)}\angle\tilde{\delta}_i^{(k)})(\overline{\tilde{I}_i^{(k)}\angle\tilde{\theta}_i^{(k)}})\rvert^2
\end{equation}
(\ref{eqn:PILoss}) is expanded to a formulation that contains only real values in the implementation so that the network training does not involve any complex arithmetics.
Combining   (\ref{eqn:PredLoss}) and (\ref{eqn:PILoss}), the loss function used to train the PIDGeuN network is
\begin{equation}
    L(\vtQ)=L_1(\vtQ)+\nu L_2(\vtQ)
\end{equation}
where $\nu$ is a factor to control the penalty on physical violation; in this study $\nu=1$ is used, which assigns both loss terms equal weights.

\subsubsection{Activation Functions} In practice, the rate of change $\dvX_g$ at the time step where a disturbance occurs can be orders of magnitude larger than those at other time steps.  Such difference was found to cause slow convergence or even divergence in the training of the PIDGeuN model, when the nonlinear activation functions in the graph MP layers are not chosen correctly.

In this study, the nonlinear activation function consists of two components.  First, a batch-normalization (BN) layer \cite{Ioffe2015} is used to reduce the potential differences in the latent vectors caused by the differences due to disturbances.  Second, the Parametric Rectified Linear Unit (PReLU) function $f(x)=\max(-\theta x,x)$ \cite{He2015} is applied to compute the new latent vectors, where $\theta>0$ is a learnable parameter.  From numerical experiments, the PReLU performed more robustly than the commonly used ReLU function, which suffered from the dying neurons problem and caused premature convergence in the training.

%% file: src/numExample.tex
\section{Numerical Examples}

\begin{figure}
    \centering
    \includegraphics[width=0.5\textwidth]{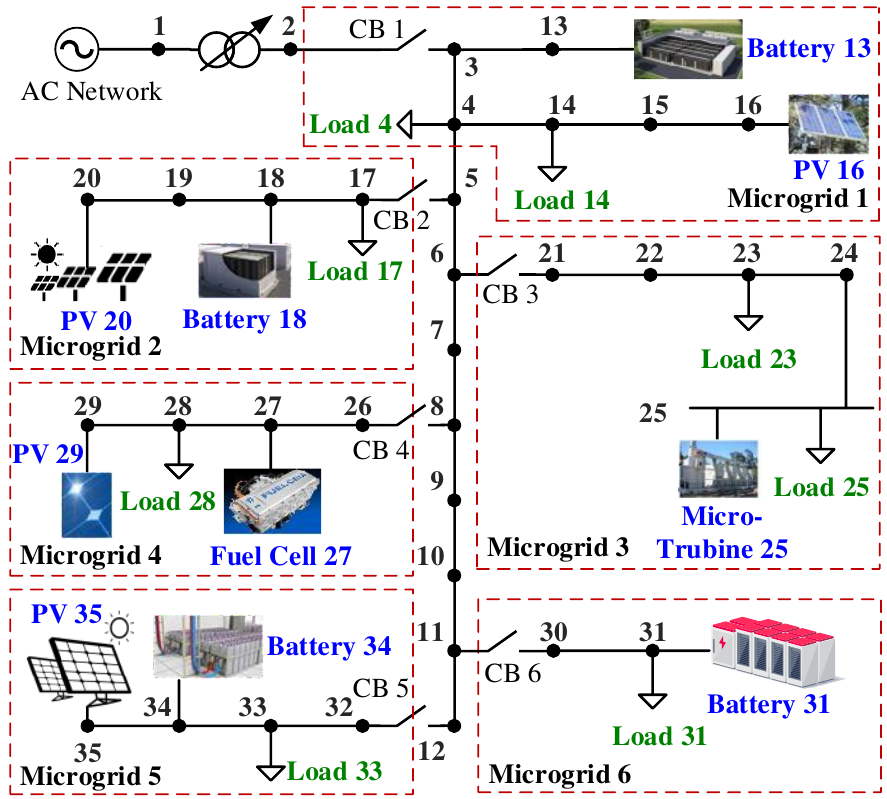}
    \caption{Networked microgrid test system.}
    \label{fig:fullGrid}
\end{figure}

In this section, the PIDGeuN architecture is applied to model and predict the transient dynamics of a typical 33-bus networked microgrid system~\cite{li2021cyber}, as shown in Fig.~\ref{fig:fullGrid}, to demonstrate its accuracy, robustness and versatility in modeling and predicting dynamics on graph. Circuit Breaker 1 is open and others are closed, so the NM system is in the islanded operation.

\subsection{Description of the Numerical Example}
\subsubsection{Datasets}  The dataset for training and testing the PIDGeuN models contains the transient responses of the  system starting from different initial conditions with a step load change applied to randomly chosen nodes at the start of simulation.  The magnitude of the load change is in the range of $\pm10\%$ of the nominal  value of each load, which is beyond the regime of linear analysis.
The DAE model of the  system is first built and simulated 
to provide synthetic data for PIDGeuN.
Two types of responses are generated: (1) \textit{complete transient response} 
that starts from an  equilibrium point and ends when the system reaches a new equilibrium point; (2) \textit{initial transient response}  during which random load changes are added every 0.01s so that the system is always away from equilibrium and shows transient dynamics.
The training dataset consists of 90 complete transient responses and initial transient responses with 5000 load changes. The test dataset contains 30 complete transient responses and initial transient responses with 500 load changes.


\subsubsection{Evaluation Metrics}  The performance of the data-driven models are quantified using three types of metrics:
\begin{compactenum}[a)]
  \item One-step root mean squared error (RMSE): The error in the rates of change of all buses is defined as
    \begin{equation}
      E_1=\sqrt{\frac{1}{TN}\sum_{k=1}^T\norm{\Tilde{\dvX}^{(k)}-\dvX^{(k)}}_F^2},
    \end{equation}
    where $T$ is the total number of time steps in the time series for prediction, and $\Tilde{\dvX}^{(k)}=\vF(\cX^{(k)},\cG)$ is the predicted rate of change using the current \textit{true} bus states.  The one-step RMSE is equivalent to the loss term in (\ref{eqn:PredLoss}) except that it is applied to the test dataset.
  \item Cumulative RMSE (C-RMSE): The difference between the predicted and true dynamics of bus states is defined as
      \begin{equation}
          E_2=\sqrt{\frac{1}{TN}\sum_{k=1}^T\norm{\tilde{\vX}^{(k)}-\vX^{(k)}}_F^2},
      \end{equation}
    where  $\tilde{\vX}^{(k)}$ is the predicted bus states at time step $k$ that are evaluated iteratively using   (\ref{eqn:nextState}) given only the initial condition $\cX^{(0)}$.  The C-RMSE accounts for the accumulation of prediction error in the time-series prediction and thus is the major metric for assessing the model performance.
    \item Number of parameters, which measures the complexity of each model. More trainable parameters give the neural network more expressive power but may result in the overfitting issue and increased computational cost for training and prediction.
\end{compactenum}

\subsubsection{Implementation Details of Nominal PIDGeuN Model}


The PIDGeuN architecture is implemented using PyTorch Geometric (PyG) \cite{PyG2019}, an open-source PyTorch-based machine learning framework for Graph Networks. 
The hyperparameters used for the nominal PIDGeuN model are: $N_M=3$, $N_A=N_C=5$, $K=5$, $D=128$, and $C=3$.
During the training, the states as well as the rate of change are normalized to a range of $[0,1]$.  
The loss is minimized using the standard Adam optimizer with an exponential decay of learning rate from $10^{-3}$ to $10^{-7}$.

\subsection{Comparison with Baseline Methods}
\subsubsection{Baseline Methods}
The PIDGeuN model is benchmarked with a number of baseline methods ranging from conventional data-driven models that do not account for graph topology to various forms of STGNN that are specialized for time series prediction.  These methods are listed as follows:
\begin{enumerate}[a)]
    \item Subspace identification~\cite{van2012subspace}: A linear state-space system identification method using only the measured states.
    \item Long Short-Term Memory (LSTM) \cite{Sutskever2014}: A type of Recurrent Neural Network (RNN) that have been widely applied for time-series prediction. We used 5 stacked layers each with a hidden size of 128.
    \item Graph Convolutional Recurrent Network (GCRN) \cite{Seo2018}: A type of STGNN that uses a Chebyshev GCN (i.e., GCL) in space and a GRU in time for data correlation over a larger spatiotemporal scale. We used 5 stacked layers each with a hidden size of 128 and $K=5$ for the GCL.
    \item Spatial-Temporal GCN (STGCN) \cite{Bing2018}: A type of STGNN that uses a GCL in space and 1D convolution instead of a RNN in time, which eliminates the usage of recurrent architecture and allows for faster training with fewer parameters. We used 2 temporal convolution layers, and 5 GCL layers with $K=5$, all with a hidden size of 128.
\end{enumerate}

The LSTM model is implemented with the PyTorch package, and the two recurrent GNN models are implemented using the PyTorch Geometric Temporal package \cite{Rozemberczki2021}.  The LSTM does not utilize the graph structure, therefore at the time step $t=t_k$, the extended node states $\vX^{(k)}\in\bR^{N\times 10}$ are stacked into a $\bR^{10N}$ vector as input to the network. 

\subsubsection{Results and Discussion}\label{sec:baseline}

In this experiment, we compare the performance of PIDGeuN against other baseline methods in the predictions of transient response of the test system.

The evaluation metrics are detailed in Table \ref{tab:comparison}, including the one-step RMSE's for the training and test dataset, and the C-RMSE's for 200 and 700 time steps.  The two C-RMSE's are chosen to quantify the short-term and long-term predictive capabilities of the models.  Note that, when compared to the complete response cases, the dynamics of the initial response is more complex due to the frequently introduced disturbances, and thus the C-RMSE of initial responses is expected to be higher than that of complete responses.

Overall, the PIDGeuN outperforms the baseline methods by a significant margin. First, all the models achieve low training and test one-step RMSE's, showing that they are sufficiently complex and expressive to predict the rate of change if given the true states, and generalize to unseen inputs.  Yet the PIDGeuN achieved the lowest training and test errors, highlighting its superior expressiveness and generalizability over other models.
Second, the high C-RMSE's show that most of the baseline models fail to produce accurate predictions over a long time horizon; particularly the STGCN quickly diverges beyond 200 time steps.  The best baseline model is in fact the subspace model, a linear method.  On the contrary, the PIDGeuN consistently achieved the lowest C-RMSE's in all cases and the slowest growth in the error, which demonstrates its robustness in time series prediction.
Finally, note that the superior performance of the PIDGeuN is achieved only using an amount of parameters that is comparable to the smallest and worst learning-based baseline model, i.e., the STGCN.


\begin{table*}
    \centering
    \caption{Comparison of PIDGeuN  and baseline models}\label{tab:comparison}
    \begin{tabular}{cccccc}    
    \hline
    \multirow{2}{*}{Model} & Training one-step & Test one-step & C-RMSE 200 & C-RMSE 700 &\multirow{2}{*}{\# of parameters} \\
     & RMSE$(\times 10^{-3})$ & RMSE$(\times 10^{-3})$ &Complete/Initial & Complete/Initial  \\\hline
     Subspace & \--- & \--- & $2.018/9.978$  & $4.815/36.791$ & \textbf{7595} \\
     LSTM & $0.167$ & $1.282$ & $12.192/304.777$ & $45.183/1348.707$ &$1830088$ \\
     GCRN &  $0.988$& $2.203$  & $6.045/26.191$   &$20.894/99.472$   & $5000726$\\
     STGCN &  $0.338$ & $3.426$  & $19.953/35.719$  &$4.886E51/1.042E52$   &$805518$\\\hline
     \textbf{PIDGeuN }   &   \textbf{0.093}&   \textbf{0.167}& \textbf{0.986$/$1.276}&   \textbf{2.311$/$11.911}&   986006\\\hline
    \end{tabular}   
\end{table*}

\begin{figure}
    \centering
    \includegraphics[width=0.5\textwidth]{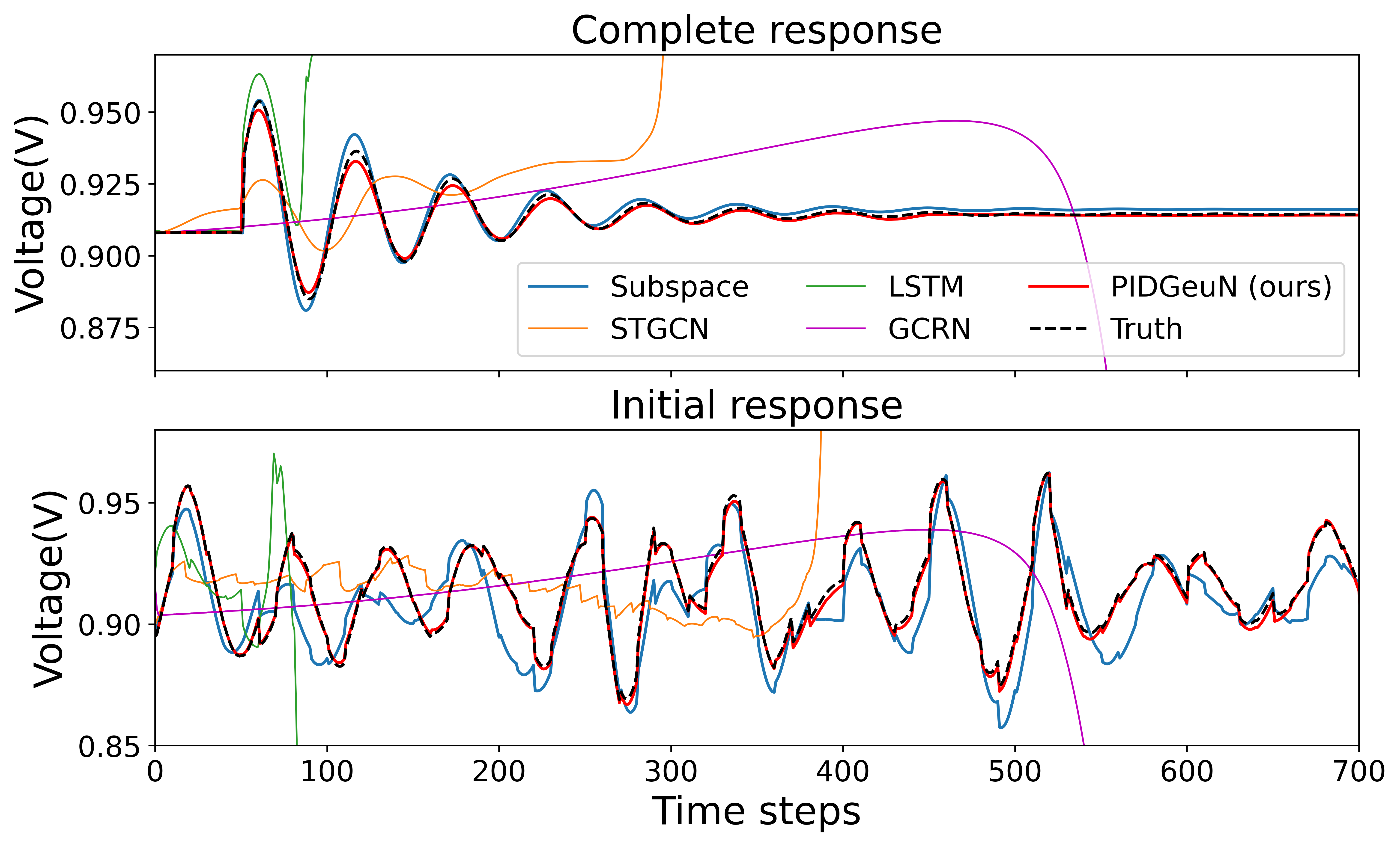}
    \caption{Complete and initial voltage dynamic response of a DER bus.}
    \label{fig:results}
\end{figure}

Subsequently, a typical voltage response of a DER bus is closely examined in Fig. \ref{fig:results}. In the complete response case, a large load change is introduced in the grid at time $t_k=50$ causing an initial step change in the DER voltage, followed by a damped oscillatory response until a new equilibrium in the system is reached.  The complete response resembles that of a linear system to a step input.  In the initial response case, smaller load changes are introduced every 10 time steps, and the system never reaches equilibrium in between the load changes.  As a result, the system dynamics in the initial response case is more dynamic and nonlinear, and thus more challenging to predict. It also mimics the applications of PIDGeuN in the real world when the system is under frequent  disturbances.

As visualized in Fig. \ref{fig:results}, in the complete response case, the PIDGeuN reproduces the DER's response to the load change by accurately predicting first the initial step voltage of the DER and then the decay of signal oscillations that matches the true dynamics in both magnitude and frequency. The subspace method performs well overall except for not capturing the peak voltages in each oscillation and the final equilibrium voltage.
The LSTM and STGCN are able to follow the first few periods of oscillation but both diverged because of the modeling error accumulation, resulting in large error. The GCRN model does not capture any oscillatory response and also diverged. In the initial response case, the PIDGeuN captures the transient dynamics almost perfectly despite the frequent introduction of load changes. The subspace method performs worse than in the previous case and misses most of the peak voltages. The other three models show similar trend as in the previous test case, and do not make any useful prediction.

From the comparison of the evaluation metrics and typical response cases, it is clear that the PIDGeuN model significantly outperforms the baseline models, including a classical system identification method (subspace), a learning-based method without graph (LSTM), and two learning-based methods with graph information (STGCN and GCRN), in terms of the generalizability, predictive accuracy, and robustness in long-term prediction.


\subsection{Ablation Study}\label{sec:ablation}
To explain the effectiveness of the PIDGeuN and study how its components affect the performance, we conducted an ablation study where a number of hyperparameters are varied one by one while holding others at the nominal value.  Specifically, the composition of the processors, the sizes of the MP layers, and the inclusion of physics-informed loss are examined.  The complete list of tested models is provided in Table \ref{tab:ablation}, where in each group the varying parameters are highlighted and the nominal model is labelled as C1. In this study, only the test case using complete response is presented for the conciseness of the paper.

\begin{table}[h!]
    \centering
    \caption{Ablation study}\label{tab:ablation}
    \begin{tabular}{m{1.2em}*{3}cm{1.2em}*{2}{m{0.6em}}cc}    
    \hline
    \multirow{2}{*}{Model} & \multirow{2}{*}{$N_A$} & \multirow{2}{*}{$N_C$} & \multirow{2}{*}{PI-loss} & \multirow{2}{*}{$K$} & \multirow{2}{*}{$D$} & \multirow{2}{*}{$C$}& C-RMSE & C-RMSE\\
    &&&&&&&200&700\\\hline
    \textbf{C1} & \textbf{5} & \textbf{5} & \textbf{True} & \textbf{5} & \textbf{128} & \textbf{3} & $0.986$ & \textbf{2.311} \\\hline
    C2 & \cg 3 &\cg  5 & True & 5 & 128 & 3 & $1.087$ & $2.634$ \\
    C3 & \cg 1 &\cg  5 & True & 5 & 128 & 3 & \textbf{0.941} & $2.825$ \\
    C4 & \cg 0 &\cg  5 & True & 5 & 128 & 3 &$2.494$ & $1.850E3$ \\
    C5 & \cg 0 &\cg  10 & True & 5 & 128 & 3 &$2.310$ & $6.231$ \\
    C6 & \cg 5 &\cg  3 & True & 5 & 128 & 3 &$1.115$ & $3.108$ \\
    C7 & \cg 5 &\cg  1 & True &  5 & 128 & 3 &$3.359$ & $6.941E61$ \\
    C8 & \cg 5 &\cg 0 & True & \--- & 128 & 3 &$3.736$ & $1.552E51$\\
    C9 & \cg 10 &\cg 0 & True  & \--- & 128 & 3 &$2.493$ & $3.591E42$\\
    C10 &\cg  3 &\cg  3 & True & 5 & 128 & 3 &$0.985$ & $2.648$\\
    C11 &\cg  1 &\cg  1 & True & 5 & 128 & 3 &$28.823$ & NaN \\\hline
    C12 & 5 & 5 &\cg  False & 5 & 128 & 3 &$1.563$ & $3.732$\\\hline
    C13 & 5 & 5 & True &\cg  7 & 128 & 3 &$1.102$ & $3.101$ \\ 
    C14 & 5 & 5 & True &\cg  3 & 128 & 3 &$1.484$ & $2.869$ \\
    C15 & 5 & 5 & True &\cg  1 & 128 & 3 &$7.385$ & $3.717E6$ \\\hline
    C16 & 5 & 5 & True & 5 &\cg  64 & 3 &$1.054$ & $2.787$ \\
    C17 & 5 & 5 & True & 5 &\cg  32 & 3 &$1.470$ & $2.614E3$ \\
    C18 & 5 & 5 & True & 5 &\cg  256 & 3 &$1.010$ & $2.489$ \\\hline
    C19 & 5 & 5 & True & 5 & 128 &\cg  1 &$7.080$ & $4.472E31$ \\
    C20 & 5 & 5 & True & 5 & 128 &\cg  2 &$1.045$ & $2.581$ \\
    C21 & 5 & 5 & True & 5 & 128 &\cg  5  &$1.338$ & $2.782$\\\hline
    
    \end{tabular} 
\end{table} 

\begin{figure}
    \centering
    \includegraphics[width=0.5\textwidth]{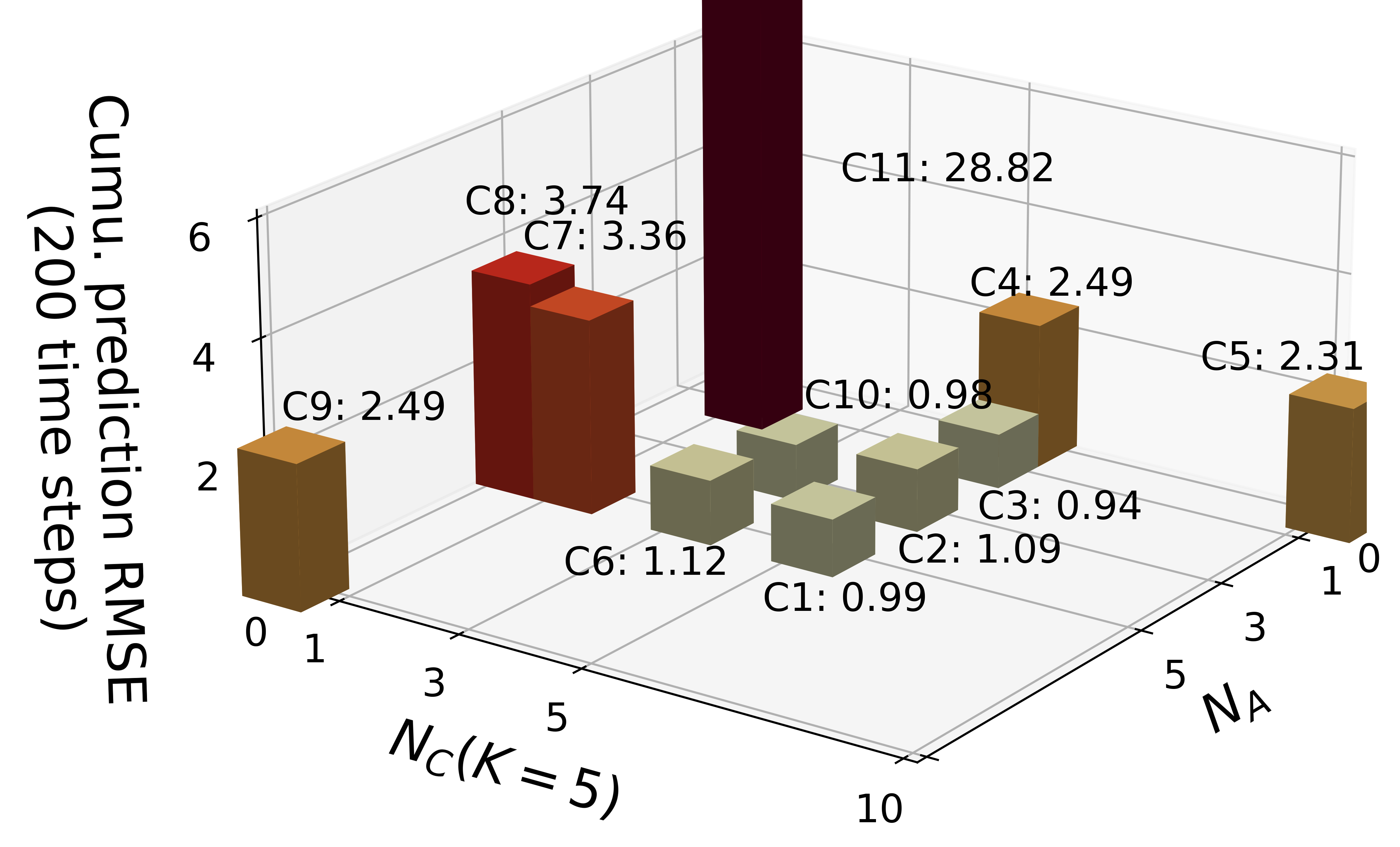}
    \caption{Impact of processor layers on the model performance.}
    \label{fig:ablation3Dbar}
\end{figure}

\subsubsection{Processor layers}


To examine how the composition of the processor affects the model performance, the models C2-C11, having different combinations of GAL and GCL layers, are trained and tested.
The differences among the models are two-fold.  First, when $N_A=0$ (GCL-only) or $N_C=0$ (GAL-only), the PIDGeuN model no longer uses a hybrid architecture; the GCL-only models (C4, C5) and GAL-only models (C8, C9) use only the physics-based and data-driven adjacency matrices, respectively.
Second, while the rest models all use a hybrid architecture, they differ significantly in the number of MP steps, which determines the capability of the network to propagate information between distant nodes, as noted in Sec. \ref{sec:mp}.

The comparison of C-RMSE's for these cases are listed in Table \ref{tab:ablation} and the C-RMSE's for 200 steps are visualized in Fig. \ref{fig:ablation3Dbar}.  
First, the nominal PIDGeuN model (C1) and the other hybrid models outperforms the models with only GCLs (C4, C5) or only GALs (C8, C9) by a significant margin.  The difference indicates the performance gain in microgrid dynamics prediction is facilitated by the proposed physics-data-infusion strategy that combines the physics-based admittance information and the data-driven attention coefficients in the adjacency matrix.
\begin{figure*}
    \centering
    \includegraphics[width=0.95\textwidth]{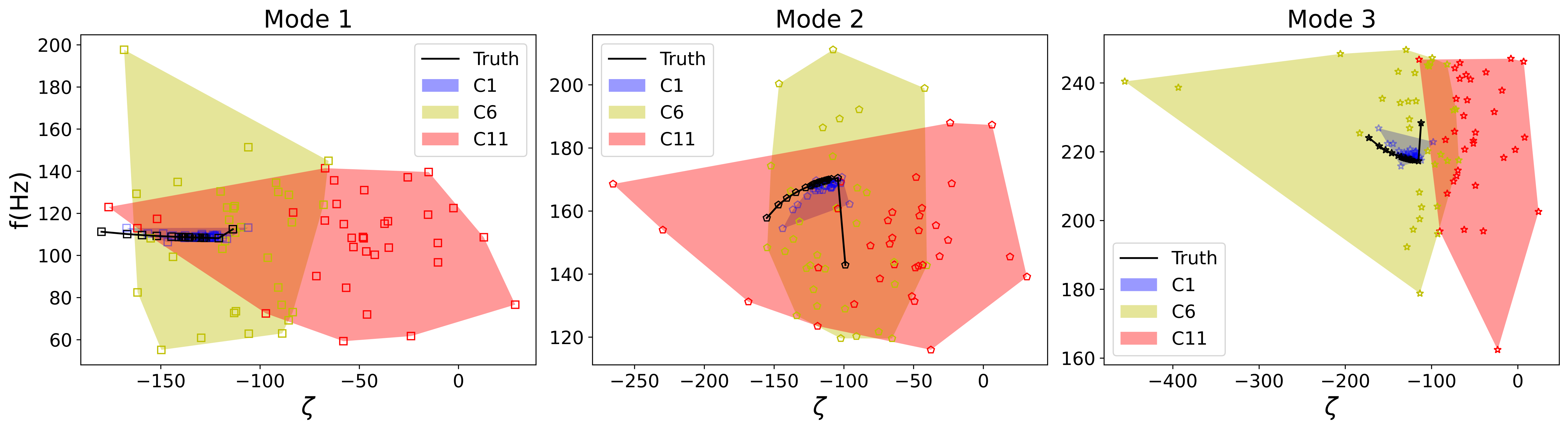}
    \caption{Frequencies and damping ratios of the first three modes in voltage dynamics, and the corresponding model predictions.}
    \label{fig:freqDamping}
\end{figure*}

Next, focusing on the models with hybrid architecture, a strong correlation between prediction accuracy and number of MP steps is identified.  
The hybrid models with relatively fewer MP steps (C4, C7-C9, C11) tend to accumulate large errors during time series prediction and diverge in some test cases.  Particularly, for models C7 and C11, the lack of MP steps limits the long-range information propagation between the nodes and the prediction performance is even worse than the GCL-only and GAL-only models.
The rest hybrid models (C1-C3, C6, C10) achieved similar performance in the 200-step prediction, with C3 being the best. But in the 700-step prediction, model C1, which has the most MP steps, consistently produced the lowest prediction error for all test cases.  The comparison indicates the importance of using sufficient number of MP steps to achieve high accuracy and robustness in the long-term prediction.

For the microgrid problem, since a load disturbance on the selected few load buses triggers a dynamical response in the whole grid, the network needs sufficient MP steps to ensure the global effects are captured. The lack of MP steps may also explain the poor performance of GCRN and STGCN in Sec. \ref{sec:baseline}. However, increasing the number of layers in GCRN and STGCN makes them vulnerable to over-smoothing issue in GNNs, and incurs prohibitive computational cost in training. 

Lastly, to better understand the gap of performance in the models, the dynamic responses are examined in further detail in terms of frequency and damping ratio, which are important from the dynamical system modeling perspective.  Three models are selected for analysis: (1) C1, the nominal and best model; (2) C6, a model that is less accurate than C1; (3) C11, a model that diverges in long-term prediction.  The frequencies $f$ and damping ratios $\zeta$, i.e., the eigenvalues, of the first three dominating oscillation modes are extracted from the voltage response using the auto-regressive moving average (ARMA) method \cite{Smail1999}, and compared against the true values that are obtained from the eigenvalue analysis of the analytical DAE model.  The results are plotted in Fig. \ref{fig:freqDamping}, where each data point corresponds to a node in the system and the shaded region illustrates the spread of the identified eigenvalues.  In the true model, most of the nodes share similar frequencies but have different damping ratios.  The eigenvalues of the PIDGeuN models differ drastically. Model C1 accurately captures most of the frequencies and damping ratios on different nodes for all three modes, which aligns with its low prediction errors. The model C6 captures some damping ratios in the first two modes and over-predicts many in the third mode; it also misses most of the frequencies.  As a result, the predictions of C6 show premature convergence to the equilibrium voltage and therefore higher error than C1. The model C11 under predicts most of the damping ratios, and shows positive damping on some nodes, which can explain its early divergence in short-term prediction in many test cases.


\subsubsection{Physics-informed Loss Function} 
\begin{figure}
    \centering
    \includegraphics[width=0.5\textwidth]{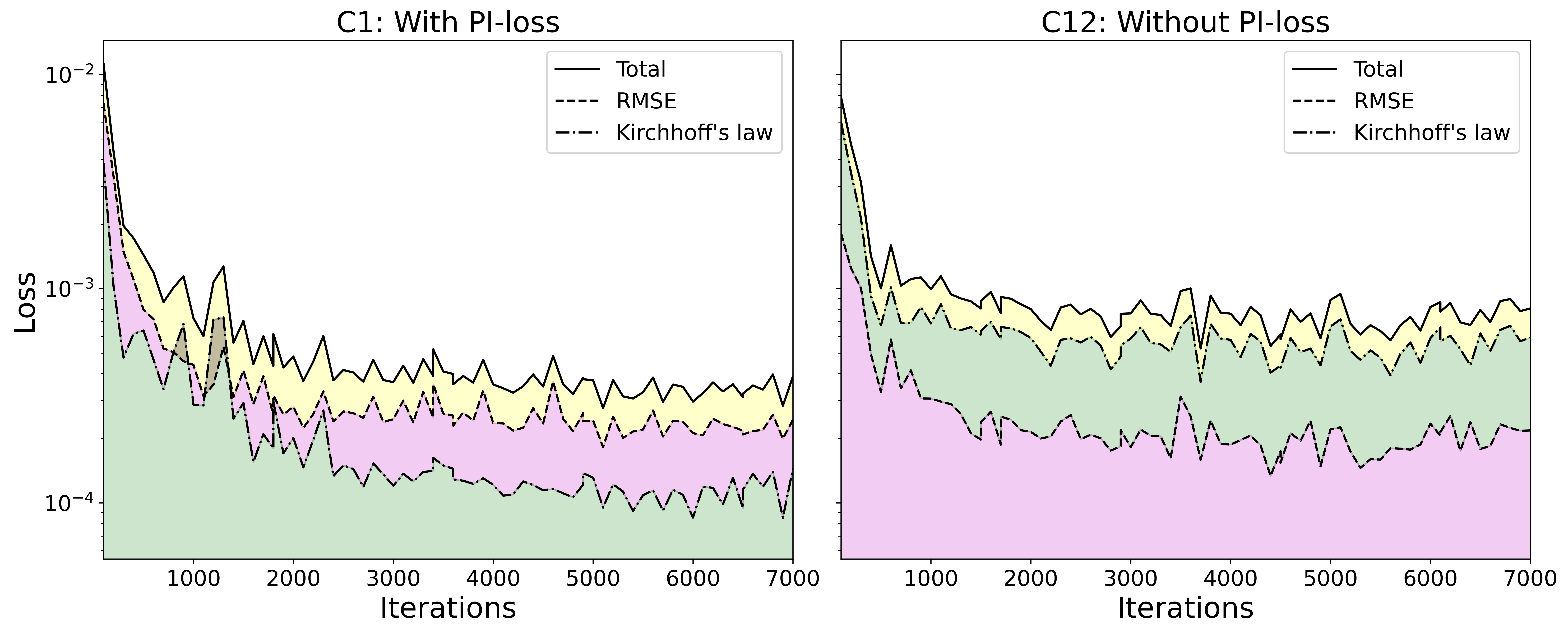}
    \caption{Comparison of loss function components in training.}
    \label{fig:loss}
\end{figure}

Next, the effect of the physics-informed loss term (\ref{eqn:PILoss}), based on the Kirchhoff's law, is examined using model C12, where the PI loss is removed during the training process.  Comparing the losses in Table \ref{tab:ablation}, it is clear that the PI loss positively contributed to model performance.  A further comparison of training loss of C1 and C12 is shown in Fig. \ref{fig:loss}. For C12, the training of network depends solely on the RMSE loss, and the additional loss term is only computed for recording purpose. During the training process, the RMSE of both cases decreased at a similar rate. However, the model prediction of C12 violates the Kirchhoff's law one order of magnitude more than that of C1, resulting in a higher total loss in training. As a result, in the actual prediction, the C12 model may produce responses that prone to violate the Kirchhoff's law, which explains its higher C-RMSE's in both 200-step and 700-step predictions than C1.  Further examination of the predicted dynamics, though not shown in paper due to space limit, reveals that C12 can only capture the first few oscillations accurately and start to diverge after around 400 steps. Mathematically speaking, the PI loss term limits the learnable parameter space where the optimizer searches during the training, and results in more feasible model prediction in the tests.

\subsubsection{Other hyperparameters} 
\begin{figure}
    \centering
    \includegraphics[width=0.5\textwidth]{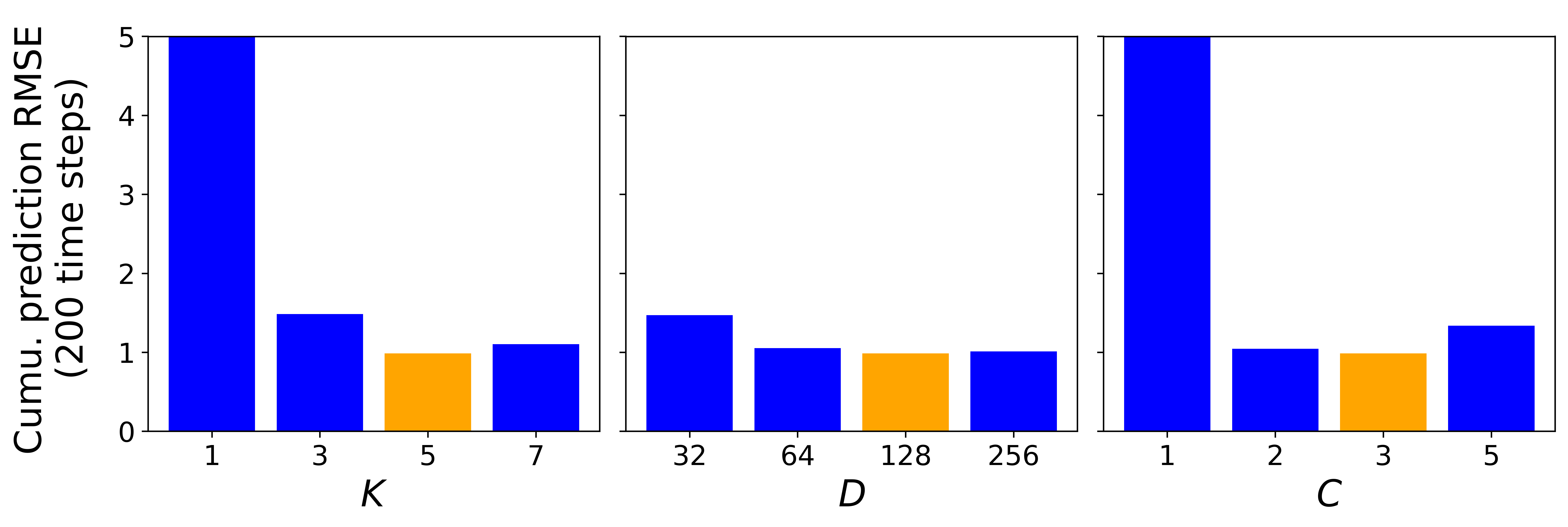}
    \caption{The effects of different hyperparameter choices on prediction performance.}
    \label{fig:ablation2Dbar}
\end{figure}

Finally, we study the effects of three hyperparameters in the PIDGeuN network through a series of models: the order of Chebyshev polynomials $K$ (C13-C15), the latent size of each hidden layer $D$ (C16-C18), and the number of steps to include in the input $C$ (C19-C21).  The results are provided in Table \ref{tab:ablation} and visualized in Fig. \ref{fig:ablation2Dbar}. In these tests, the number of GALs and GCLs are kept the same as C1, i.e., $N_A=N_G=5$. 

As found earlier, a sufficient number of MP steps is critical in the microgrid prediction problem to ensure long-range information propagation and achieve high predictive accuracy.
An economic way to increase the MP steps is to increase the polynomial order $K$ in the GCLs.
In the parametric study, increasing $K$ from $1$ to $5$ rapidly decreases the prediction error, as expected.  However, increasing $K$ beyond $3$ has a marginal improvement on the prediction performance, and when $K=7$ the C-RMSE even increased possibly due to overfitting.  The trend indicates that, once sufficient MP is reached, keep increasing $K$ does not benefit the prediction performance much, and only increases computational effort.

The size of hidden layer $D$ determines the number of trainable parameters and thus the size of a network.  The performance of PIDGeuN turns out to be less sensitive to $D$ than other hyperparameters, and $D=128$ achieves a good balance between network size and prediction accuracy.


The number of input steps $C$ decides the amount of previous information the network can access when predicting the future step. 
When $C=1$, the network only has access to the measurable bus states at the current step, which is insufficient to reconstruct  the DER controllers that dominate the dynamics. As a result the prediction performance is poor. The performance is improved immediately when another step of states is included in the input ($C=2$), but the improvement becomes marginal as more steps of previous states are included in the input. This is likely due to the nature of the current microgrid problem where a long-term temporal dependency is not significant, and two steps of measured states already form sufficient time delay embedding to fully describe the system.

%% file: src/conclusion.tex
\section{Conclusion}
In this paper, we presented the Physics-Informed Dynamic Graph Neural Network, PIDGeuN, for accurate, efficient and robust prediction of transient dynamics in microgrids.  The PIDGeuN model exploits its graph-based architecture to incorporate the topological information of  microgrids.  Furthermore, based on a judiciously designed message passing mechanism, the PIDGeuN incorporates two physics-informed techniques to improve its predictive performance. First, the PIDGeuN dynamically learns and adjusts the underlying graph representation of the system by combining the data-driven attention-based weights and physics-informed admittance-based weights, and thus better represents the inter-dependencies between buses.  Second, the PIDGeuN includes the known equation of physical law of the power system 
in the loss function that ensure the feasibility of the predictions.

The PIDGeuN is demonstrated using transient response data of microgrids due to load changes that contain complete transient responses, and initial transient responses. The results show that the PIDGeuN can accurately and robustly predict the dynamics of the microgrid using initial states and load changes in the system, and outperforms a number of baseline methods in the transient predictions.  Specifically, in the complete response cases, the PIDGeuN accurately captures the frequencies and damping ratios of the system as well as the new equilibrium states after the system stabilizes. In the initial response cases, the PIDGeuN is capable of capturing the transient and nonlinear dynamics due to the frequent load changes.  The physics-informed techniques are proven to significantly contribute to the predictive accuracy of the model.

The results establish initial capability of the PIDGeuN to be applied to large scale networked microgrids, and show its potential as an online predictive tool to enable predictive or preventive control in real time applications, which is crucial to the stable operations of the NMs. 